# Identification of main factors affecting trust and determination of their importance in electronic businesses in Iran

Mozhdeh Sadighi[1], Mohammad Mahdi Ghobadi[2], Seyyed Hossein Hasanpour Matikolaee[3]

[1]*Department of Information Technology Management, University of Tehran, Tehran, Iran*

[2]*Department of Computer Engineering, Mazandaran University of Science and Technology, Babol, Iran*

[3]*Department of Computer Engineering, Islamic Azad University, Science and Research Branch, Amol, Iran*

[1] Mozhdeh.Sadighi@ut.ac.ir
[2] mm.qobadi@gmail.com
[3] hosseinhasanpour@live.com

*Abstract*: **Today, trust has become one of the main concerns of the electronic business in Iran. The role of trust especially in electronic businesses those directly deal with selling physical goods through internet is a lot more evident. Reviewing literature shows that several factors affect establishing of trust in potential customers. Since trust establishment needs to be noticed in each triple stages of an electronic purchase (before, during and finally after purchase). In this study by using field research, the importance of influential factors affecting the potential customers in three stages of an electronic purchase is determined. Based on the results from conducting the research, the certainty of traceability of the purchase with importance factor of 85.97% in pre-purchase stage, safety of transactions and the time of delivery of goods with the importance factor of 85.67% in the middle stage of the purchase and receiving a fault-free and undamaged good with the importance factor of 89.55% in the post purchase stage make up the top three most important factors.**

## I. INTRODUCTION

The Development and growth of internet and web in recent years in Iran has had a fast and significant pace. Through increased access of people to the internet, this growth has specifically been high in business to customer (B2C) electronic business. However despite of the growth in electronic business in Iran, the number and volume of internet based transactions in this country is an insignificant amount compared to the majority of countries in the world. In recent years some activities has been conducted in order to introduce and validate electronic business. Amongst those are the Iran Festival of Web and e-namad which is an electronic symbol of trust for electronic shops.

Iran Festival of Web is one of the most credible festivals in the field of information technology in the country which is held annually and introduces best web sites in the past years. This festival by creating an atmosphere of a healthy competency among the growing Iranian websites, results in the growth and development of Iranian websites. In addition, the electronic symbol of trust (e-namad) is an indication on the behalf of the Electronic Commerce Development Center to those electronic markets that adhere to a series of rules and regulations legislated by this center. This symbol acts like an Iran Standard Logo which is issued for physical goods with this difference that the e-namad is issued solely for electronic shops and can be viewed in every compliant shops website. The goal is to gain the customers trust in purchasing from online stores and adhering to the fact of recognizing the rights of consumers. The major goal of activities like this is to promote the use of internet, stimulating the virtual business and building trust in the community towards online shopping. Despite all of the actions taken so far, public confidence in internet based commerce is low. The importance of this matter is especially higher in case of physical goods and expensive goods. Although there are successful cases such as Digikala, Takhfifan, Axprint and others which managed to gain their customers trust very well, these are sparse in numbers.

In today's commerce world, customer orientation is one of the most important factors of success in businesses. Trust encourages customers to have a purchase from a store and if satisfied, repeat the process. The more the trust is established, the more people risk taking is increased. Thus gaining customer trust by different methods is an act of recognizing and meeting consumer needs. Trust is a variable that depends on many factors, knowing these factors and their importance is an important step toward helping the business managers in order to increase their customers and guarantee their existence in the fast changing market. The goal of this study is to determine the influential factors on customer trust on businesses that have electronic shops. Identifying and addressing these factors will let managers turn their potential customers into active and loyal customers.

This research focuses on factors affecting the purchase of physical goods via internet. Physical goods are goods that can be touched such as cell phones, CD, garment, food, etc. Since it is not possible to touch and see these kinds of goods, purchasing these kinds of goods are more sensitive that electronic crafts. Therefore, the role of trust establishment for a customer regarding these kinds of goods is more important in comparison with digital products and services. Since some of trust building factors and their importance are different in potential and practical customers, it is worthy to note that by customers in this study we mean practical customers. Therefor based on the explanations that have been stated so far, the general questions of this research are as follows:

1. What are the influential factors on customer trust toward electronic businesses?
2. Which level does each of these factors belong to?
3. How much is the importance of factors affecting customer trust (confidence) at every stage of the purchase?

The structure of this research is as follows: First a review on the literature is done, then, the suggested framework is presented. In the next step, the method in which the research is based on and the statistical population and sample is explained. The proceeding section includes the findings. References and summery form the last sections of this research.

## II. LITERATURE REVIEW

Trust is a key component in the marketing literature and a requirement in electronic commerce [1]. So far so many different definitions for trust have been proposed by researchers. The various definitions had two reasons. First, the abstract nature of the concept is sometimes considered equivalent with other concepts such as validity, reliability or certainty. Moreover, being a multi-dimensional aspect causes different dimensions (cognitive, emotional, and behavioral factors) to be included [2]. Thus, in various fields trust has been studied in different ways.

Trust refers to a set of beliefs that the online customer has about his electronic supplier and his possible behavior in future [3]. Kim and others (2009) have defined trust as a multidimensional concept that based on purpose can be related to a company, customer, website, product, etc. Trust in cyberspace is the most important factor that makes the customer accepts the risk of transaction [4]. Trust is an important factor in any relationship that the trustee (e.g. customer) doesn't have direct control over the actions of the trustor (e.g. dealer or store), the decision is important and the environment is uncertain as well [1]. Trust is a concept that allows a person to make decision in a situation where there is uncertainty or risk [5].

As it is evident in many definitions, trust is always accompanied by risks or uncertainty. In fact in situations where clients do not have a complete understanding of the seller or do not have a former experience in purchasing from him, they are made to risk. In such circumstances, there are several factors that can gain customer trust and make them risk. Moorman et al. (1997) present a model for trust-building and its consequences. They believe that there are five antecedents which can lead to users trust. These are "Individual user characteristics", "perceived researcher interpersonal characteristics", "perceived user organizational characteristics", "perceived inter-organizational/interdepartmental characteristics" and "perceived project characteristics" [6].

They Lee and Turban (2001) stated that four factors affect customer trust in internet shopping. These factors include the reliability of the sellers, reliability of internet as a shopping means, contextual or infrastructural factors and others (such as firm size and demographical variables). Egger in 2001 presented a trust model for electronic commerce. This model consists of four components which include the pre-interactive filters that happen before online interaction, web interface features, website information content and relationship management. This model fully covers the buyer-seller interaction [7]. Kim and others (2005) presented a multidimensional model for trust in B2C e-commerce transactions. In order to present this framework they first described the structure and components of the ecommerce. They believe that the transaction process in internet is done among four different entities, including the buyer, seller, third party and the technology. Therefor they offered six dimensions of the technology, behavioral, organizational, information, products and transaction and their respective sub-dimensions to describe the structure of online trust [8].

As it has been observed so far different terms (such as components, antecedents, requirements, dimensions, determinants and principles) concerning trust has been used in different researcher's works. Therefore, the definition of the term trust is sometimes due to disagreement among researchers in the field are used interchangeably. Also many studies only have sufficed describing a model or a framework for trust in online space and did not specify the importance of each factor on the Trust. In this study, the factors affecting customer trust when purchasing from an electronic stores in Iran is studied.

Table 1 shows a summary of the literature review.

*Proposed framework*

In proposed framework the purchase process is taken into consideration. Many studies that have been done, have examined the role of trust before purchase. Some of them also paid attention to the security issues during shopping process. This study by breaking the electronic shopping process into three distinct stages of pre-purchase, mid-purchase and post-purchase, studies trust in each of these stages. The affecting factors that are listed in the following framework are obtained by utilizing the literature and then validating by expert. Since the questions of the experts validation questionnaire was designed in an open format, they were asked to mention factors that seem important to them and also factors which are not addressed in the questionnaire. Six factors were identified and added to the framework. So the final framework is localized for Iran (Table 2). In order to distinguish added factors, they are shown in Italic font style.

TABLE I
FACTORS INFLUENCING CUSTOMER TRUST IN INTERNET SHOPPING

| Factor | Reference |
|---|---|
| Brand, credit and reputation of the organization | [1], [9], [10], [8], [7], [11], [12], [13], [14] |
| The quality of the clients previous experience of the internet and online purchases | [8], [10], [15], [7], [13] |
| Feedback quality of other customers | [8], [7], [11], [13] |
| Appearance of the website (issues related to graphic design) | [16], [17], [2], [18], [8], [7], [11], [19], [13] |
| Creditability of the partners (e.g. payment gateways, investors, infrastructures supplier, third-party assurance warranty and guarantee companies, etc.) | [8], [10], [7], [11], [19], [13] |
| Traceability of purchase (ensuring post-purchase support) | [7] |
| Instant guidance (such as online chat, phone, etc.) | [7], [11] |
| The transparency, accuracy, being updated and completeness of the information of the site, including: 1) Organizational information 2) product information (such as price and quality) 3) Statistics (e.g. number of members, visitors, and so on) | [16], [20], [18], [8], [7], [11], [19] |
| Infrastructure (such as the technical performance of site) | [18], [10], [21], [8], [7] |
| Security in transactions such as using SSL to ensure the privacy of personal information, card information, etc. | [22], [23], [18], [8], [24], [10], [25], [15], [7], [11], [19], [12], [13], [14] |
| Offering purchase receipt | [7], [19] |
| Ability to track status of purchase online | [7], [11], [19] |
| Correct delivery (the ordered good is the delivered good) | [15] |
| Guarantee, warranty and RMA | [8], [7], [19] |
| Support (guidance, advice, technical assistance and so forth in using the commodity) | [8], [7] |
| Product quality | [8], [10], [12] |
| Incentives (e.g. bonuses, coupons, etc.) | [8], [11], [14] |
| Trust and loyalty programs (e.g. looking for customers feedbacks, welcoming their suggestions or claims, etc.) | [11] |

TABLE 2
PROPOSED FRAMEWORK BASED ON PURCHASING DECISION PROCESS

| Pre-purchase | Mid-purchase | Post-purchase |
|---|---|---|
| Brand, credit and reputation the organization | Infrastructure (such as the technical performance of site) | Ability to track status of purchase online |
| *Having e-namd symbol* | *Delivery time* | *Product condition (being fault-free, undamaged and perfect)* |
| The quality of the clients previous experience of the internet and online purchases | *Shipping method (post, freight company, etc.)* | *On time delivery* |
| Feedback quality of other customers | Security in transactions such as using SSL to ensure the privacy of personal information, card information, etc. | Correct delivery ( the ordered good is the delivered good ) |
| Appearance of the website (issues related to graphic design) | *Payment method (online, after delivery, etc.)* | Support (guidance, advice, technical assistance and so forth in using the commodity ) |
| Creditability of the partners (e.g. payment gateways, investors, infrastructures supplier, warranty and guarantee companies, etc.) | Offering purchase receipt | Incentives (e.g. bonuses, coupons, etc.) |
| Traceability of purchase (ensuring post-purchase support) | | Trust and loyalty programs |
| Instant guidance (such as online chat, phone, etc.) | | Guarantee, warranty and RMA |
| The transparency, accuracy, being updated and completeness of the information of the site , including:<br>1) Organizational information<br>2) Product information (such as price and quality)<br>3) Statistics (e.g. number of members, visitors, and so on) | | *Commodity packaging* |
| | | Product quality |

The purpose of building trust in customers in post-purchase phase is to create loyalty in them. Loyal customers are invaluable asset for businesses. Today, the strategy of many businesses is to gain customers satisfaction and trust (even at the cost of profit lost) in order to make them as their loyal and committed customers.

III. RESEARCH METHOD

The method which is employed in this study is considered as applied research from the perspective of its purpose. In terms of its primary data gathering method, it is considered as survey research that is a kind of descriptive research. First, by using the literature review secondary data were collected and the basic questionnaire was designed.

A small number of suggestions were submitted in this sections which were similar to those trust making factors already addressed in questionnaire. Therefore, no new or different factor was observed.

A statistical population of this research is the members of Mazandaran University of Science and Technology student forum who had at least one online shopping in the last year. The forum has 5056 members. Since knowing the exact number of members who had an online purchase in the last year was not possible, random sampling was done among them. In order to do so, an online questioner was designed and placed on the forum. According to the Krejcie and Morgan sample size table, for a 5000 community, 357 samples are needed [26]. After twenty days from the initial date of placing the questionnaire on the forum, the numbers of filled questioner were reached 357. Answers were stored in an SQL database and then moved into Excel software. In order to do statistical analysis on the collected data, SPSS16.0 was used.

This questionnaire was submitted to five experts for validation and some changes had been done. For example factors such as e-namad, shipping method, delivery time and so forth were suggested by the experts and added to the framework.

The final questionnaire consists of 36 questions and phrase. The primary section which belongs to the personal information of respondents includes eight questions. The second section consists of questions that are used to evaluate the importance of each influential factors affecting trust in each stage of electronic shopping. In the last section, the respondents were asked to write down any factor that they deem important but has been neglected in the questionnaire.

*A. Validity*

The validity of tools determines that to what extent the designed tool evaluated the intended concept. By validity of the questionnaire, questioner wants to understand that if the research questionnaire evaluates what it needs to evaluate exactly or the questions asks something else that is not intended. For evaluating the questionnaire validity, three experts in the field of e-commerce who were university professors were consulted.

*B. Reliability*

Reliability implies that if a lot of sampling is done, research results remains stable and different results will not obtained. Usually, in order to make sure about questionnaire reliability, Cronbach's Alfa coefficient is used. Because of the complexity of calculating this coefficient, researchers use SPSS software to calculate it. By doing so, if the Cronbach's Alfa coefficient is greater than 0.7, the questionnaire can be said to be reliable. In this study, the Cronbach's Alfa coefficient was estimated 0.894 which denotes that the questionnaire has a fine reliability.

## IV. FINDINGS

*a) Descriptive statistics*

Among the 357 samples, 67.2% of the responders were female and 32.8 % were men. Four ranges were defined for age. According to these ranges, 46.3 % of the responders aged 20 to 25 years, 35.8 % aged 25 to 30 years, 11.9 % were 30 to 35 years old and 6 % had more than 35 years old. The highest age in the study sample was 42 years.

Among the responders, 56.7% had a bachelor's degree, 35.8 % had a master's degree, 6 % had a doctoral degree, and 1.5 % had a high school diploma. The field of study of the majority of responders, equal to 59.7 %, was computer software engineering, 17.9 % information technology, 9 % management, 7.5 % Electrical Engineering and 6 % industrial engineering. The number of purchases via Internet for about 43 % of responders in a recent year was more than six times.

*b) Inferential statistics*

For 40.3% of responders, the level of satisfaction with purchase in the statistical sample of this study was high and for 37.3% was average. In a question, responders were asked to rate their trust level towards electronic shopping in the form of Liker spectrum ranging from very low to very high. The results show that the trust level of 44.8% of them is average and 17.9% is low.

As presented it Table 3, the results show that the importance of trust in all three stages of electronic purchase has a high importance. Based on the answers of the statistical sample, the order of the importance of these stages with very minor difference is as follows: mid-purchase 0.86%, pre-purchase 0.85% and post-purchase 0.83%. Prioritizing the importance of each of these influential factors affecting trust in each stage, according to the opinions of the participants in this study is shown separately in the following table.

TABLE 3
PRIORITIZATION OF THE IMPORTANCE OF FACTORS AFFECTING TRUST

| Pre-purchase | | Mid-purchase | | Post-purchase | |
|---|---|---|---|---|---|
| **Factor** | **importance** | **Factor** | **importance** | **Factor** | **importance** |
| 1) Traceability of purchase | 85.97% | 1-1) Safety in transaction<br>1-2) *Delivery time* | 85.67% | *1) Product condition* | 89.55% |
| 2) Brand, credit and reputation the organization | 81.49% | *2) Payment method* | 83.28% | 2-1) Correct delivery<br>2-1) Product quality | 89.25% |
| 3-1) The quality of the clients previous experience of the internet and online purchases<br>3-2) Site information | 78.51% | *3) Shipping method* | 80.00% | 3) Guarantee, warranty and RMA | 87.46% |
| 4) Instant guidance | 76.72% | 4) Offering purchase receipt | 79.40% | *4) On time delivery* | 85.37% |
| 5) Creditability of partners | 74.63% | 5) Infrastructure | 74.93% | 5) Ability to track the status of purchase online | 83.28% |
| 6) Feedback quality of other customers | 73.73% | | | 6) Support | 81.79% |
| *7) Having e-namad symbol* | 72.84% | | | 7) Trust and loyalty programs | 71.34% |
| 8) Site appearance | 71.34% | | | *8) Commodity packaging* | 69.55% |
| | | | | 9) Incentives | 63.58% |

## V. CONCLUSION AND DISCUSSION

Trust that is the customer expectation about seller's motivations and behaviors is a necessary component in each relationship such as buyer-seller relationship in businesses [8]. The main goal of this article is to identify main influential factors affecting Iranian customers trust towards shopping from internet-based stores. In order to achieve this goal, a framework using literature review is designed and then by the help of experts gotten revised.

In this research trust has been surveyed in three stages of pre-purchase, mid-purchase and post-purchase. The importance of trust and the methods of building trust are different in these stages. Therefor businesses should employ different strategies for each stage. Using the questionnaire tool, the viewpoints of 357 responders are collected. Results (Table 3) show that out of 25 factors, 23 of them have an importance of more than 71% and the other two have a factor above 63%. As a result, it can be claimed that the identified influential factors affecting trust establishment in physical goods shopping from online stores are attributable to a high degree.

In association with the six factors added by experts, it is conveyed by the results that although these factors are not regarded well in literature, from the responders' point of view they have high importance. These factors include specifying payment methods, product delivery method and the time needed for product delivery by the seller during the purchase and also the product status in terms of being fault-free in after purchase stage. Therefore, it is evident that in association with the four aforementioned factors, respondents' satisfaction in their previous internet based shopping experiences were low and it led to the reduction of their trust level. It can be concluded that whilst the aforementioned factors are the necessities of any electronic purchase in the world, but there are still significant problems around satisfying these necessities in Iran.

In pre-purchase phase, results indicate that although many existing factors in final framework are already parts of legal requirements of e-namad, it has a low importance in pre-purchase stage compared to other influential factors affecting trust. This shows that although it has been some time that the e-namad project is being conducted by government in the country, lots of people still do not know much about it and its application or to them this symbol is not a guarantee of trust and reliability of the store complying with it. Therefore, it is suggested that in future researches, e-namad is studied thoroughly and the reasons for such a low effect on customers trust is identified. The problems pertaining to its implementation and execution studied and solutions are presented.

## ACKNOWLEDGMENT

The authors would like to thank all those who participated in this study.

## REFERENCES

[1] S. L. Jarvenpaa, N. Tractinsky and M. Vitale, "Consumer trust in an internet store," *Information Technology and Management,* pp. 45-71, 2000.

[2] Y. D. Wang and H. H. Emurian, "An overview of online trust: Concepts,elements, and implications," *Computers in Human Behavior,* vol. 21, pp. 105-125, 2005.

[3] K. Coulter and R. Coulter, "Determinants of Trust in a Service Provider: the Moderating Role of Length of Relationship," *Journal of Services Marketing,* vol. 16, pp. 35-50, 2002.

[4] D. J. Kim, D. L. Ferrin and H. R. Rao, "Trust and Satisfaction, Two Stepping Stones for Successful E-Commerce Relationships: A Longitudinal Exploration," *Information Systems Research,* vol. 20, no. 2, pp. 237-257, 2009.

[5] N. Luhmann, "Familiarity, Confidence, Trust: Problems and Alternatives," in *Trust: Making and Breaking Cooperative Relations*, Oxford, University of Oxford, 2000, pp. 94-107.

[6] C. Moorman, R. Deshpande and G. Zaltman, "Factors affecting trust in market research relationship," *Journal of Marketing,* vol. 57, pp. 81-101, 1993.

[7] F. N. Egger, "Affective Design of E-Commerce User Interfaces:How to Maximise Perceived Trustworthiness," in *The International Conference on Affective Human Factors Design*, London, 2001.

[8] D. J. Kim, Y. I. Song, S. B. Braynov and H. R. Rao, "A multidimensional trust formation model in B-to-C e-commerce:a conceptual framework and content analyses of academia/practitioner perspectives," *Decision Support Systems,* pp. 143-165, 2005.

[9] p. M. Doney and J. P. Cannon, "An examination of the nature of trust in buyer-seller relationships," *Journal of Marketing,* vol. 61, pp. 35-51, 1997.

[10] M. K. O. Lee and E. Turban, "A Trust Model for Consumer Internet Shopping," *International Journal of Electronic Commerce,* pp. 75-91, 2001.

[11] A. Mukherjee and P. Nath, "Role of electronic trust in online retailing: A re-examination of the commitment-trust theory," *European Journal of Marketing,* vol. 41, pp. 1173-1202, 2007.

[12] M. S. Velmurgan, "SECURITY AND TRUST IN E-BUSINESS: PROBLEMS AND PROSPECTS," *International Journal of Electronic Business Management,,* vol. 7, pp. 151-158, 2009.

[13] J. He, "The vital role of trust in E-commerce: A meta-analysis," *International Journal of E-Business Development,* vol. 3, no. 3, pp. 97-107, 2013.

[14] A. Bryant and B. Colledge, "Trust in Electronic Commerce Business Relationships," *Journal of Electronic Commerce Research,* Vols. 32-39, 2002.

[15] D. L. Hoffman, T. P. Novak and M. Peralta, "Building consumer trust online," *Communications of the ACM,* vol. 42, pp. 80-85, 1999.

[16] M. C. Roy, O. Dewit and B. A. Aubert, "The impact of interface usability on trust in Web retailers," *Internet Research,* vol. 11, no. 5, pp. 388-398, 2001.

[17] D. Cyr, "Modeling Web Site Design Across Cultures: Relationships to Trust, Satisfaction, and E-Loyalty," *Journal of Management Information Systems,* vol. 24, no. 4, pp. 47-72, 2008.

[18] M. I. Eid, "Determinants of e-commerce customer satisfaction, trust, and loyalty in Saudi Arabia," *Journal of Electronic Commerce Research,* vol. 12, pp. 78-93, 2011.

[19] S. Srinivasan, "Role of trust in e-business success," *Information Management & Computer Security,* vol. 12, pp. 66-72, 2004.

[20] D. H. McKnight, V. Choudhury and C. Kacmar, "Developing and Validating Trust Measures for e-Commerce: An Integrative Typology," *Information Systems Research,* vol. 13, pp. 334-359, 2002.

[21] J. Lee and N. Moray, "Trust, control strategies, and allocation of functions in human-machine systems," *Ergonomics,* vol. 35, pp. 1243-1270, 1992.

[22] C. Flavián and M. Guinalíu, "Consumer trust, perceived security and privacy policy: Three basic elements of loyalty to a web site," *Industrial Management & Data Systems,* vol. 106, no. 5, pp. 601-620, 2006.

[23] T. B. Warrington, N. j. Abgrab and H. M. Caldwell, "Building trust to develop competitive advantage in e-business relationships," *Competitiveness Review,* vol. 10, no. 2, pp. 160-168, 2000.

[24] D. W. Manchala, "E-Commerce trust metrics and models," *IEEE INTERNET COMPUTING,* vol. 4, no. 2, pp. 36-44, 2000.

[25] H. Hsiung, S. Scheurich and F. Ferrante, "Bridging E-Business and Added Trust: Keys to E-Business Growth," *IT professional,* vol. 3, no. 2, pp. 41-45, 2001.

[26] R. V. Krejcie and D. W. Morgan, "DETERMINING SAMPLE SIZE FOR RESEARCH ACTIVITIES," *EDUCATIONAL AND PSYCHOLOGICAL MEASUREMENT,* pp. 607-610, 1970.

[27] B. W. Wirtz and N. Lihotzky, "Customer retention management in the B2C electronic business," *Long Range Planning,* pp. 517-532, 2003.